\documentclass{jetpl}
\twocolumn
\title{Evolution of edge states in topological superfluids during the quantum phase transition}

\rtitle{Evolution of edge states in topological superfluids}

\sodtitle{Evolution of edge states in topological superfluids}

\author
{M.A. Silaev$^{*}$ \/\thanks{e-mail: msilaev@mail.ru,volovik@boojum.hut.fi}
and G.E. Volovik $^{+\#}$
}

\rauthor{
M.A. Silaev, G.E.Volovik
}

\sodauthor{
Silaev, Volovik
}

\address{
$^{*}$ Institute for Physics of Microstructures RAS, 603950 Nizhny Novgorod, Russia\\
$^{+}$ Low Temperature Laboratory, Aalto University, School of Science and
Technology, P.O. Box 15100, FI-00076 AALTO, Finland
\\
$^{\#}$ Landau Institute for Theoretical Physics RAS, Kosygina 2, 119334 Moscow, Russia
}

\abstract{ The quantum phase transition between topological and
non-topological insulators or between fully gapped
superfluids/superconductors can occur without closing the gap. We
consider the evolution of the Majorana edge states on the surface
of topological superconductor during transition to the
topologically trivial superconductor on example of non-interacting
Hamiltonian describing the spin-triplet superfluid  $^3$He-B. In
conventional situation when the gap is nullified at the
transition, the spectrum of Majorana fermions shrinks and vanishes
after the transition to the trivial state. If the topological
transition occurs without the gap closing, the Majorana fermion
spectrum disappears by escaping to ultraviolet, where Green's
function approaches zero.  This demonstrates the close connection
between the topological transition without closing the gap and
zeroes in the Green's function. Similar connection takes place in
interacting systems where zeroes may occur due to interaction. }

\begin{document}

\maketitle


\section{Introduction}

General properties of fermionic spectrum in condensed matter and particle physics
are determined by topology of the ground state (vacuum). The classification schemes based on topology \cite{Schnyder2008,Schnyder2009a,Schnyder2009b,Kitaev2009,Volovik2003,Volovik2007,Horava2005} suggest in particular the classes of  topological insulators and fully gapped topological superfluids/superconductors.  The main signature of such topologically nontrivial vacua with the energy gap in bulk
is the existence of zero-energy edge states on the boundary or at the interface between topologically distinct
domains \cite{HasanKane2010,Xiao-LiangQi2011}.  In Refs. \cite{Volovik2003,Volovik2007,Horava2005} the classification is based on topological properties of matrix Green's function, while the other schemes explore the properties of single
particle Hamiltonian and thus are applicable only to systems of free (non-interacting) fermions.
As was found in Ref.  \cite{FidkowskiKitaev2010}, classifications of interacting and non-interacting fermionic systems do not necessarily coincide. This is related to zeroes of the Green's function, which according to Ref.  \cite{Volovik2007} contribute to topology alongside with the poles. Due to zeroes the integer topological charge of the interacting system can be changed without closing the energy gap, and it is suggested that this may lead to the occurrence of topological insulators with no fermion zero modes on the  interface \cite{Gurarie2011,Essin2011}.

 \begin{figure}[top]
\centerline{\includegraphics[width=0.9\linewidth]{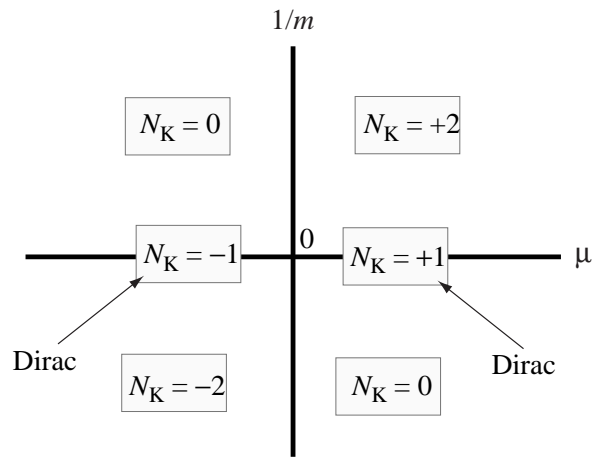}}
  \caption{\label{QPT}  Fig.1. Phase diagram of topological states of triplet superfluid of $^3$He-B type in
  equation \eqref{eq:Hamiltonian} in the plane $(\mu,1/m)$. States on the line
  $1/m=0$ correspond to the  Dirac vacua, which Hamiltonian is non-compact. Topological charge of the Dirac fermions  is intermediate between charges of compact $^3$He-B states.
   The line $\mu=0$ marks topological quantum phase transition, which occurs between the weak coupling $^3$He-B  (with $\mu>0$, $m>0$ and topological charge $N_K=2$) and the strong coupling $^3$He-B   (with $\mu<0$, $m>0$ and $N_K=0$).    This transition is topologically equivalent to quantum phase transition between Dirac vacua with opposite mass parameter
 $M=\pm |\mu|$.  The gap in the spectrum becomes zero at this transition.
The line $1/m=0$ separates the states with different asymptotic behavior of the Hamiltonian at infinity:
$H({\bf p}) \rightarrow \pm \tau_3 p^2/2m$. The transition across this line  occurs without closing the gap,
 }
\end{figure}

In principle, the analogous situation with zero in the Green's function and quantum phase transition without closing the energy gap may also occur in the free fermion case. Example is provided by superfluid $^3$He-B, which belongs to the topologically nontrivial class
of fully gapped superfluids, which possesses Andreev-Majorana fermions on the surface
\cite{SalomaaVolovik1988,CCGMP,Davis2008,Nagai2008,Murakawa2009c,ChungZhang2009,Volovik2009a,Nagato2009,Tsutsumi2011}. In the phase diagram of the topological superfluid/superconductor
of  the $^3$He-B type in Fig. \ref{QPT}, the topological quantum phase transition (TQPT) between two states with different topological charges across the line $1/m=0$ occurs without closing the gap \cite{Volovik2009b}.  Instead, the asymptotic behavior changes at the momentum infinity, at $p\rightarrow \infty$, where the Hamiltonian diverges and thus the Green's function approaches zero.  Such scenario is impossible in the models with the bounded Hamiltonian \cite{Gurarie2011,Essin2011}, which takes place  in approximation of finite number of crystal bands. We consider the evolution of the spectrum of Majorana fermions on the surface of a topological superfluid when the system crosses the lines of the TQPT with and without closing the gap.

 \section{Spectrum of edge states}

 The invariant $N_K$ relevant  for $^3$He-B in Fig. \ref{QPT} is the  topological invariant protected by symmetry:
\begin{equation}
N_K = {e_{ijk}\over{24\pi^2}} ~
{\bf tr}\left[K \int   d^3p
~H^{-1}\partial_{p_i}H
H^{-1}\partial_{p_j}H
H^{-1}\partial_{p_k}H
\right]\,,
\label{3DTopInvariant_tau}
\end{equation}
where $K$ is matrix which  commutes or anti-commutes with the Hamiltonian.
The proper model Hamiltonian which has the same topological properties as superfluids/superconductors of  the $^3$He-B class is the following :
\begin{equation}
H=\left(\frac{p^2}{2m}-\mu\right) \tau_3-
   c \tau_1 {\mbox{\boldmath$\sigma$}} \cdot{\bf p} \,,
\label{eq:Hamiltonian}
\end{equation}
where $\tau_i$ and $\sigma_i$ are Pauli matrices of Bogolyubov-Nambu spin  and nuclear spin correspondingly; the parameter $c$ serves as the speed of light for the Dirac Hamiltonian obtained in the limit $1/m=0$ and further will shall use $c=1$. The symmetry $K$, which enters the topological invariant $N_K$ in \eqref{3DTopInvariant_tau}, is represented by the $\tau_2$ matrix, which anti-commutes with the Hamiltonian: it is combination of time reversal  and particle-hole symmetries in $^3$He-B.

Let us consider the Majorana fermions using the simplest model of
the interface between the superfluid and  the vacuum, in which the
Hamiltonian changes abruptly at the boundary
\cite{ChungZhang2009}. The boundary is at $z=0$ and at $z<0$ we
have the equation $H \psi=E\psi$ with $H=H_0+H_1$ where
 \begin{equation}\label{H0}
 H_0=\left(\frac{p^2}{2m}-\mu\right)\tau_3+\tau_1\sigma_zp_z
 \end{equation}
 \begin{equation}\label{H1}
 H_1=\tau_1(p_x\sigma_x+p_y\sigma_y)\,,
 \end{equation}
 and we use the boundary conditions $\psi (z=0)=0$. Without loss of
 generality we set $p_y=0$ so that $p_x=p_\perp=\sqrt{p^2-p_z^2}$
 and
  \begin{equation}\label{H1-1}
 H_1=\tau_1p_\perp\sigma_x.
 \end{equation}
Now it is possible to simplify the equation by choosing the wave
function transformation $\tilde{U}=\hat\sigma_z U$ and
 $\tilde{V}= V$, where $\psi=(U,V)^T_\tau$.
The Hamiltonian then transforms as
 \begin{equation}\label{H0-2}
 H_0=\left(\frac{p^2}{2m}-\mu\right)\tau_3+\tau_1p_z
 \end{equation}
  \begin{equation}\label{H1-2}
 H_1=-\tau_2p_\perp\sigma_y.
 \end{equation}
Since $\sigma_y$ becomes the good quantum number,  this representation allows to
reduce the general problem from $4\bigotimes 4$ to $2\bigotimes 2$
matrices.

Let us consider first the solutions corresponding to $\hat
\sigma_y\psi =\psi$. Then we get the following equation in the
form
$$
\left(%
\begin{array}{cc}
  \left(\frac{p^2}{2m}-\mu\right)-\varepsilon & p_z+ip_\perp \\
  p_z-ip_\perp & -\left(\frac{p^2}{2m}-\mu\right)-\varepsilon \\
\end{array}\right) \left(U\atop V\right)=0
$$
which yields the relation between $U$ and $V$:
$$
V=\frac{p_z-i p_\perp}{\varepsilon+\frac{p^2}{2m}-\mu}U
$$
and
$$
\frac{p^2_{1,2}}{2m}-\mu=-m\pm \sqrt{m^2+\varepsilon^2-2m\mu}.
$$
The solution is a superposition of two modes decaying at $z<0$
with $Im(p_{z1,2})>0$ where $p_{z1,2}^2=p_{1,2}^2-p_\perp^2$:
\begin{equation}
\psi=A \psi_1+B\psi_2\,.
\label{superposition}
\end{equation}
The boundary condition yields the equation
\begin{equation}\label{Eq:Energy}
 \frac{p_{z1}-i
p_\perp}{\varepsilon+\frac{p^2_1}{2m}-\mu}=\frac{p_{z2}-i
p_\perp}{\varepsilon+\frac{p^2_2}{2m}-\mu}.
\end{equation}

\begin{figure}[hbt]
\centerline{\includegraphics[width=0.90\linewidth]{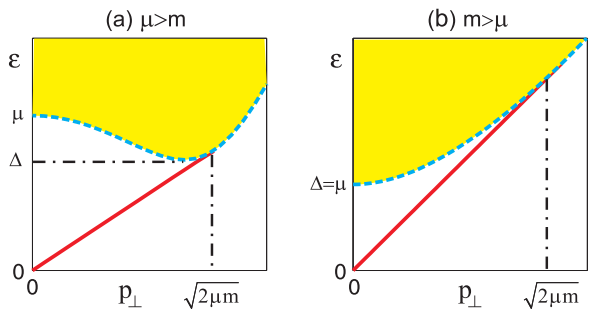}}
\caption{\label{Fig:Spectrum} Fig.~2. Spectrum of Andreev-Majorana
fermions, localized states on the surface of topological
superfluid/superconductor of the $^3$He-B class (red solid lines)
for (a) $\mu>m>0$ and (b) $m>\mu$. The spectrum of bound states
terminates when it merges with continuous spectrum in bulk (yellow
color), whose border is shown by blue dashed line.  }
\end{figure}

For $p_\perp^2<2m\mu$ this equation \eqref{Eq:Energy} has an exact
solution
\begin{equation}\label{Eq:ExactSpectrum}
\varepsilon=p_\perp\,.
\end{equation}
The wave functions $\psi_1(z)$ an $\psi_2(z)$ forming the bound
state have the localization lengths determined by equation
\begin{equation}\label{Eq:pz}
 p_{z1,2}=i\left(m \mp \sqrt{p_\perp^2+m^2-2m\mu}\right)\,.
\end{equation}
The solutions corresponding to $\hat \sigma_y\psi =-\psi$ yield
the spectrum $\varepsilon=-p_\perp$, and taking into account the
$p_y$ dependence one obtains the helical spectrum of Majorana
fermions with $H_{\rm Majorana}=c(\sigma_y p_x - \sigma_x p_y)$
\cite{ChungZhang2009}.

For $m>0$ the spectrum of Andreev-Majorana fermions
$\varepsilon=\pm p_\perp$ is shown by the red solid line in Fig.
\ref{Fig:Spectrum} for $\epsilon>0$.
The bound states are confined to the
region $|p_\perp|<\sqrt{2m\mu}$. They disappear when their
spectrum merges with the continuous spectrum in bulk. The edge of
continuous spectrum is shown by blue dashed line in
Fig. \ref{Fig:Spectrum}. If $m>\mu$  the minimum of the bulk energy spectrum
 increases
monotonically with momentum $p_\perp$, therefore the bulk gap is
\begin{equation}
\Delta=\mu~~,~~m>\mu\,.
\label{gap1}
\end{equation}
If $\mu>m$ the minimum of the bulk energy is non-monotonic function of
$p_\perp$ having the minimum at
$p^{min}_\perp=\sqrt{2m(\mu-m)}$ where the bulk gap is
\begin{equation}
\Delta=\sqrt{2\mu m-m^2}~~,~~0<m<\mu\,.
\label{gap2}
\end{equation}
The line $m=\mu$ marks the non-topological quantum phase transition -- the momentum space analog of the Higgs transition, when the Mexican hat  potential as function of $p_\perp$ emerges for $\mu > m$ \cite{Volovik2007}.

 \section{Evolution of edge state at topological quantum phase transition}

Let us first consider the behavior of the spectrum of Majorana fermions at the topological transition at which $m$ crosses zero. When $m$ approaches zero,
$m\rightarrow 0$, the region of momenta where bound states exist shrinks  and finally for $m<0$, i.e. in the topologically trivial superfluid, no bound states exist any more.
Simultaneously the gap in bulk, which at small $m$ is $\Delta\approx \sqrt{2m\mu}$ according to Eq. \eqref{gap2}, decreases with decreasing $m$
and nullifies at $m=0$. This corresponds to the conventional scenario of the topological quantum phase transition, when at the phase boundary between the two gapped states with different topological
numbers the gap is closed. The same happens at the TQPT occurring when $\mu$ crosses zero
(see phase diagram in Fig. \ref{QPT}).

Now let us consider what happens with bound states in the case if the TQPT occurs in the opposite limit, when $m$ changes sign via infinity, i.e. when $1/m$ crosses zero. This topological transition occurs without  closing of the gap. In this case the bound states formally exist for all $p_x$ even in the limit $1/m \rightarrow 0$. However, in this limit the ultraviolet divergence takes place.
Two components in the superposition \eqref{superposition}
for the wave function of Majorana fermion have characteristic lengths determined by imaginary momentum in Eq. \eqref{Eq:pz}.
At $1/m\rightarrow 0$ these lengths become
\begin{eqnarray}
 \nonumber
   &L^{-1}_{z1}={\rm Im}~ p_{z1}\approx\mu-p_\perp^2/(2m)\rightarrow \mu \,,\\
   &L^{-1}_{z2}={\rm Im}~ p_{z2}\approx 2m \rightarrow \infty \,.
\end{eqnarray}
The length of the first component remains finite in this limit but the
dimension of the second component shrinks to zero and thus leaves the region of applicability of the
model Hamiltonian. As a result the wave function of the bound state cannot be constructed any more.
 In other words, if the TQPT from topologically non-trivial to the trivial  insulator (or superconductor) occurs without closing the gap, the gapless spectrum of surface states disappears by escaping via ultraviolet.

\section{Conclusion}

We considered two scenarios of the evolution of the spectrum of the gapless edge states at TQPT.
One scenario refers to the traditional case, when the gap is nullified at the transition. In this case, when the TQPT from the topological state to the topologically trivial one is approached the spectrum of Majorana fermions shrinks to zero and vanishes after the transition.

The other scenario takes place, when the TQPT occurs without the gap closing.
In this case, the  spectrum of Majorana fermions vanishes by escaping to ultraviolet. The characteristic momentum of one of the wave functions relevant for the forming of the bound state diverges at  the TQPT as ${\rm Im}~ p_{z2}=2m \rightarrow \infty$. This limit corresponds to formation of zero of the Green's function, $G=1/(i\omega -H)\rightarrow 0$. Thus, similar
to the interacting systems,  the two phenomena -- TQPT without the gap closing and zeroes in the Green's function -- are closely related.  That is why we expect that the same scenario with escape to the ultraviolet takes place for the interacting systems:  if due to zeroes in the Green's function the TQPT in bulk occurs without closing the gap, the spectrum of edge states will nevertheless change at the TQPT, and  this change occurs via the ultraviolet.

In future it will be interesting to extend the consideration to the bulk-vortex correspondence
\cite{Volovik1999,Nishida2010,SilaevVolovik2010,TeoKane2010b,Herbut2010}, i.e.
to study the evolution of the spectrum of fermion zero modes
localized inside the core of vortices (or other topological defects) in the process of topological quantum phase transition in bulk with and without the gap closing. The escape via the ultraviolet can be also considered for the gapless topological matter with the fermion zero modes forming flat bands
\cite{Ryu2002,SchnyderRyu2010,HeikkilaVolovik2011,HeikkilaKopninVolovik2011}
 and Fermi arcs  \cite{Tsutsumi2011,XiangangWan2011,Burkov2011}.

\section*{\hspace*{-4.5mm}ACKNOWLEDGMENTS}
It is a pleasure to thank C. Kane and A. Ludwig for helpful
discussion on zeroes in Green's function. This work is supported
in part by the Academy of Finland and its COE program 2006--2011,
by the Russian Foundation for Basic Research (grants 09-02-00573-a
and 11-02-00891-à), and by the Program ``Quantum Physics of
Condensed Matter'' of the Russian Academy of Sciences, ``Dynasty''
foundation and Presidential RSS Council (grant MK-4211.2011.2).



\begin{thebibliography}{99}

\bibitem{Schnyder2008}
A.P. Schnyder, S. Ryu, A. Furusaki and A.W.W. Ludwig,
Classification of topological insulators and superconductors in three spatial dimensions,
Phys. Rev. {\bf B~ 78}, 195125 (2008).

\bibitem{Schnyder2009a}
A.P. Schnyder, S. Ryu, A. Furusaki and A.W.W. Ludwig,
Classification of topological insulators and superconductors,
 AIP Conf. Proc. {\bf 1134}, 10 (2009);    arXiv:0905.2029.

\bibitem{Schnyder2009b}
A.P. Schnyder, S. Ryu and A.W.W. Ludwig,
Lattice model of three-dimensional topological singlet superconductor with time-reversal symmetry
Phys. Rev. Lett. {\bf 102}, 196804 (2009);
arXiv:0901.1343.

\bibitem{Kitaev2009} A. Kitaev,
Periodic table for topological insulators and superconductors,
AIP Conference Proceedings, Volume {\bf 1134}, pp. 22-30 (2009);
  arXiv:0901.2686.

\bibitem{Volovik2003} G.E. Volovik, {\it The Universe in a Helium Droplet}, Clarendon
Press,  Oxford (2003), http://ltl.tkk.fi/personnel/THEORY/volovik/book.pdf

 \bibitem{Volovik2007} G.E. Volovik,
 Quantum phase transitions from topology in momentum space,
 in:  "Quantum Analogues: From Phase Transitions to Black Holes and Cosmology",
 eds.  W.G. Unruh and R. Sch\"utzhold,
 Springer Lecture Notes in Physics {\bf 718} (2007), pp. 31--73;
cond-mat/0601372.


\bibitem{Horava2005}  P. Ho\v{r}ava,
Stability of Fermi surfaces and $K$-theory,
Phys. Rev. Lett. \textbf{95}, 016405 (2005).

\bibitem{HasanKane2010}
M.Z. Hasan and C.L. Kane,
Topological Insulators,  Rev. Mod. Phys. \textbf{82}, 3045 (2010).

\bibitem{Xiao-LiangQi2011}
Xiao-Liang Qi and Shou-Cheng Zhang,
Topological insulators and superconductors,
Rev. Mod. Phys. {\bf 83}, 1057--1110 (2011).


\bibitem{FidkowskiKitaev2010}
L. Fidkowski and A. Kitaev,
Effects of interactions on the topological classification of free fermion systems.
Phys. Rev. B {\bf 81}, 134509 (2010);
 Topological phases of fermions in one dimension,
 Phys. Rev. B {\bf 83}, 075103 (2011).

\bibitem{Gurarie2011}
V. Gurarie,
Single-particle GreenÕs functions and interacting topological insulators,
Phys. Rev. B {\bf 83}, 085426 (2011).

\bibitem{Essin2011}
A.M. Essin and V. Gurarie,
Bulk-boundary correspondence of topological insulators from their respective GreenÕs functions,
Phys. Rev. B {\bf 84}, 125132 (2011).


\bibitem{SalomaaVolovik1988}
 M.M. Salomaa and  G.E. Volovik,
 Cosmiclike domain walls in superfluid $^3$He-B: Instantons and diabolical points in (${\bf k}$,
${\bf r}$) space," Phys. Rev.  {\bf B~37}, 9298--9311 (1988).

\bibitem{CCGMP} C.A.M. Castelijns, K.F. Coates, A.M. Gu\'enault, S.G. Mussett and G.R. Pickett,
Landau critical velocity for a macroscopic object moving in superfluid $^3$He-B: evidence for
gap suppression at a moving surface,
Phys. Rev. Lett.  {\bf 56}, 69--72 (1986).

\bibitem{Davis2008}
J.P. Davis, J. Pollanen, H. Choi, J.A. Sauls, W.P. Halperin and A.B. Vorontsov,
Anomalous attenuation of transverse sound in $^3$He,
Phys. Rev. Lett. {\bf 101}, 085301 (2008).

\bibitem{Nagai2008}
K. Nagai,  Y. Nagato, M. Yamamoto and S. Higashitani,
Surface bound states in superfluid $^3$He,
J.  Phys.  Soc.  Jap. {\bf 77},   111003 (2008).



  \bibitem{Murakawa2009c}
S. Murakawa, Y. Tamura, Y. Wada, M. Wasai, M. Saitoh, Y. Aoki, R.
Nomura, Y. Okuda, Y. Nagato, M. Yamamoto, S. Higashitani and K. Nagai,
New anomaly in transverse acoustic impedance of superfluid
$^3$He-B with a wall coated by several layers of $^4$He,
 Phys. Rev. Lett. {\bf 103}, 155301 (2009).

\bibitem{ChungZhang2009}
Suk Bum Chung, Shou-Cheng Zhang,
Detecting the Majorana fermion surface state of $^3$He-B through spin relaxation,
Phys. Rev. Lett. {\bf 103}, 235301 (2009);
arXiv:0907.4394.


\bibitem{Volovik2009a}
 G.E. Volovik,
Fermion zero modes at the boundary of superfluid $^3$He-B,
 Pis'ma ZhETF {\bf 90}, 440--442 (2009); JETP Lett. {\bf 90}, 398--401 (2009);
arXiv:0907.5389.



\bibitem{Nagato2009}   Y. Nagato, S. Higashitani and K. Nagai,
   Strong anisotropy in spin suceptibility of superfluid He-3-B film caused by surface bound states,
   J. Phys. Soc. Japan    {\bf 78}, 123603  (2009).

\bibitem{Tsutsumi2011}
Y. Tsutsumi, M. Ichioka and K. Machida,
Majorana surface states of superfluid $^3$He A and B phases in a slab,
Phys. Rev. B {\bf 83}, 094510 (2011).

\bibitem{Volovik2009b}
G.E. Volovik,
Topological invariant  for superfluid  $^3$He-B and quantum phase transitions,
Pis'ma ZhETF {\bf 90}, 639--643 (2009);  JETP Lett. {\bf 90}, 587--591 (2009);
arXiv:0909.3084.

\bibitem{Volovik1999}
G.E. Volovik,
Fermion zero modes on vortices in  chiral superconductors",
Pis'ma ZhETF {\bf 70} 601--606 (1999);  JETP Lett. {\bf 70}, 609-614 (1999);
cond-mat/9909426.

\bibitem{Nishida2010}
Y. Nishida,
Is a color superconductor topological?
Phys. Rev. D \textbf{81}, 074004 (2010).

\bibitem{SilaevVolovik2010}
M.A. Silaev and G.E. Volovik,
Topological superfluid $^3$He-B: fermion zero modes on interfaces and in the vortex core, J. Low Temp. Phys. {\bf 161},  460--473 (2010);
arXiv:1005.4672.

\bibitem{TeoKane2010b}
J.C.Y. Teo  and  C.L. Kane,
Topological defects and gapless modes in insulators and superconductors,
Phys. Rev. B {\bf 82}, 115120 (2010).

\bibitem{Herbut2010}
Chi-Ken Lu and I.F. Herbut,
Pairing symmetry and vortex zero-mode for superconducting Dirac fermions,
Phys. Rev. B {\bf 82}, 144505 (2010)

\bibitem{Ryu2002}
S. Ryu and  Y. Hatsugai,
Topological origin of zero-energy edge states in particle-hole symmetric systems,
 Phys. Rev. Lett. {\bf 89}, 077002 (2002).

 \bibitem{SchnyderRyu2010}
 A.P.  Schnyder and S. Ryu,
Topological phases and flat surface bands in superconductors without inversion symmetry,
 arXiv:1011.1438;
Phys. Rev. B {\bf 84}, 060504(R) (2011).

 \bibitem{HeikkilaVolovik2011}
T.T. Heikkil\"a and G.E. Volovik,
Dimensional crossover in topological matter: Evolution of the multiple Dirac point in the layered system to the flat band on the surface,
Pis'ma ZhETF {\bf 93}, 63--68 (2011); JETP Lett. {\bf 93}, 59--65 (2011);
arXiv:1011.4185.

 \bibitem{HeikkilaKopninVolovik2011}
T.T. Heikkil\"a, N.B. Kopnin and G.E. Volovik,
Flat bands in topological media,
Pis'ma ZhETF {\bf 94}, 252-- 258 (2011); JETP Lett. {\bf 94}, 233--239(2011);
 arXiv:1012.0905.

\bibitem{XiangangWan2011}
Xiangang Wan, A.M. Turner,  A. Vishwanath  and S.Y. Savrasov,
 Topological semimetal and Fermi-arc surface states in the electronic structure of pyrochlore iridates,
Phys. Rev. B {\bf 83}, 205101 (2011).

\bibitem{Burkov2011}
A.A. Burkov and L. Balents,
Weyl semimetal in a topological insulator multilayer,
Phys. Rev. Lett. {\bf 107}, 127205 (2011).



\end{thebibliography}
\end{document}